# Machine-Learning-Guided Polymorph Selection in Molecular Beam Epitaxy of $In_2Se_3$

Running title: Polymorph selection of molecular beam epitaxy grown $In_2Se_3$ via Bayesian Optimization

Running Authors: Trice et al.

Ryan Trice[1], Mingyu Yu[2], Eric Welp[3], Morgan Applegate[1], Wesley Reinhart[1], Stephanie Law[1,3,4,5, a)]

[1]Department of Materials Science and Engineering, Pennsylvania State University, University Park, PA 16802, USA
[2]Department of Materials Science and Engineering, University of Delaware, 201 Dupont Hall, 127 The Green, Newark, Delaware 19716, United States
[3]Department of Physics, The Pennsylvania State University, University Park, Pennsylvania 16802-6300, USA
[4]Materials Research Institute, Pennsylvania State University, University Park, PA 16802, USA
[5]2D Crystal Consortium Material Innovation Platform, The Pennsylvania State University, University Park, Pennsylvania 16802

a) Electronic mail: sal6149@psu.edu

Indium selenide ($In_2Se_3$), a layered chalcogenide with multiple polymorphs, is a promising material for optoelectronic and ferroelectric applications. However, achieving polymorph-pure thin films remains a major challenge due to the complex growth space. In this work, Bayesian Optimization (BO) is successfully leveraged to guide the molecular beam epitaxy (MBE) growth of $In_2Se_3$ on $Al_2O_3$ substrates. By training a predictive Gaussian Process Regressor with sequential learning, we efficiently explored substrate temperature, indium flux, selenium flux, and cracker temperature, reducing experimental trials required for successful synthesis. A $\gamma$-$In_2Se_3$ film with 91% phase purity was achieved in fewer than ten BO run samples. Attempts to isolate $\alpha$-$In_2Se_3$ were limited by amorphous film formation at low temperatures, indicating that single-step co-deposition is unsuitable for crystalline $\alpha$-$In_2Se_3$ on $Al_2O_3$. Overall, this study validates Bayesian Optimization as a powerful approach for phase-selective growth in complex materials systems.

## I. INTRODUCTION

Synthesis of high-quality crystalline thin films is often done using techniques like molecular beam epitaxy (MBE), chemical vapor deposition (CVD), and pulsed laser deposition (PLD). These methods offer precise control over film thickness, composition, and crystallinity by tuning growth parameters such as material flux or flow rate, material flux ratios, substrate temperature, chamber pressure, etc. [1–5]. While this tunability enables tailored synthesis, it also introduces significant complexity, resulting in multidimensional growth spaces. Identifying optimal growth parameters within these complex spaces typically involves trial-and-error experimentation, in which growers use a combination of intuition and grid search methods to guide the choice of growth parameters. These trial-and-error approaches are often time-consuming, costly, and heavily reliant on user expertise [6–9]. Grid search methods also require exponentially more experiments as the number of synthesis parameters increases[7,10]. Design of experiments (DoE) techniques have often been used to help reduce the number of needed experiments and provide a way to find the optimal growth parameters [9,11]. While more efficient than grid searching, DoE still requires many experiments to adequately sample the space and find the global optimal set of growth parameters. Additionally, DoE assumes that the response surface is relatively smooth and can be modeled with predetermined low-order polynomials, which may not hold true in complex materials systems[11,12].

An alternative way to navigate complex growth spaces is through the use of Bayesian Optimization (BO), which can explore parameter space more efficiently and find global optimal growth parameters[8,11,13–18]. By leveraging probabilistic models such as Gaussian Process Regression (GPR), BO can statistically select the most informative experiments, balancing exploration of new growth parameter spaces with exploitation of known good parameters. This approach accelerates the discovery of optimal synthesis conditions, reduces experimental overhead, and improves reproducibility[8,11,13–16].

In thin film synthesis, BO has been used to optimize the epitaxial growth rate of Si[19], achieve high residual resistivity ratios in $SrRuO_3$ [20,21], improve the thermoelectric properties of Sn doped $In_{1-x}Ga_xAs_{1-y}Sb_y$[22], and increase the photoluminescence intensity of $WS_2$[23]. Beyond BO, many other machine learning (ML) techniques are being employed to better understand synthesis-structure-property relationships in thin films. These approaches discover correlations between synthesis conditions and characterization data, including reflection high-energy electron diffraction (RHEED) patterns, X-ray diffraction (XRD) peak positions and full width at half maximum (FWHM), Raman spectra peak positions and FWHM, and atomic force microscopy (AFM) images [24–30]. ML has thus contributed significantly to understanding the interplay between growth parameters and fundamental material characteristics for complex material systems.

One material system that has been underexplored with BO and ML techniques is the indium-selenium (In-Se) family, which contains the layered chalcogenide material $In_2Se_3$. $In_2Se_3$ exhibits multiple polymorphs (including α, β, γ, and κ), each with distinct structural and electronic properties[31,32]. These polymorphs are promising for applications in optoelectronics[33–37], phase-change memory[38–42], and ferroelectric devices[43–47]. Due to their distinct properties, leveraging $In_2Se_3$ for devices requires polymorph-pure films. The crystal structures of the different polymorphs can be seen in Figure 1. In addition to the multiple polymorphs, the In–Se phase diagram includes several competing phases, such as InSe and $In_4Se_3$, which must be avoided during growth[48]. Achieving phase-pure and polymorph-pure $In_2Se_3$ is further complicated by the sensitivity of polymorph formation to growth parameters like temperature, In to Se flux ratios, and substrate material[31,49–57].

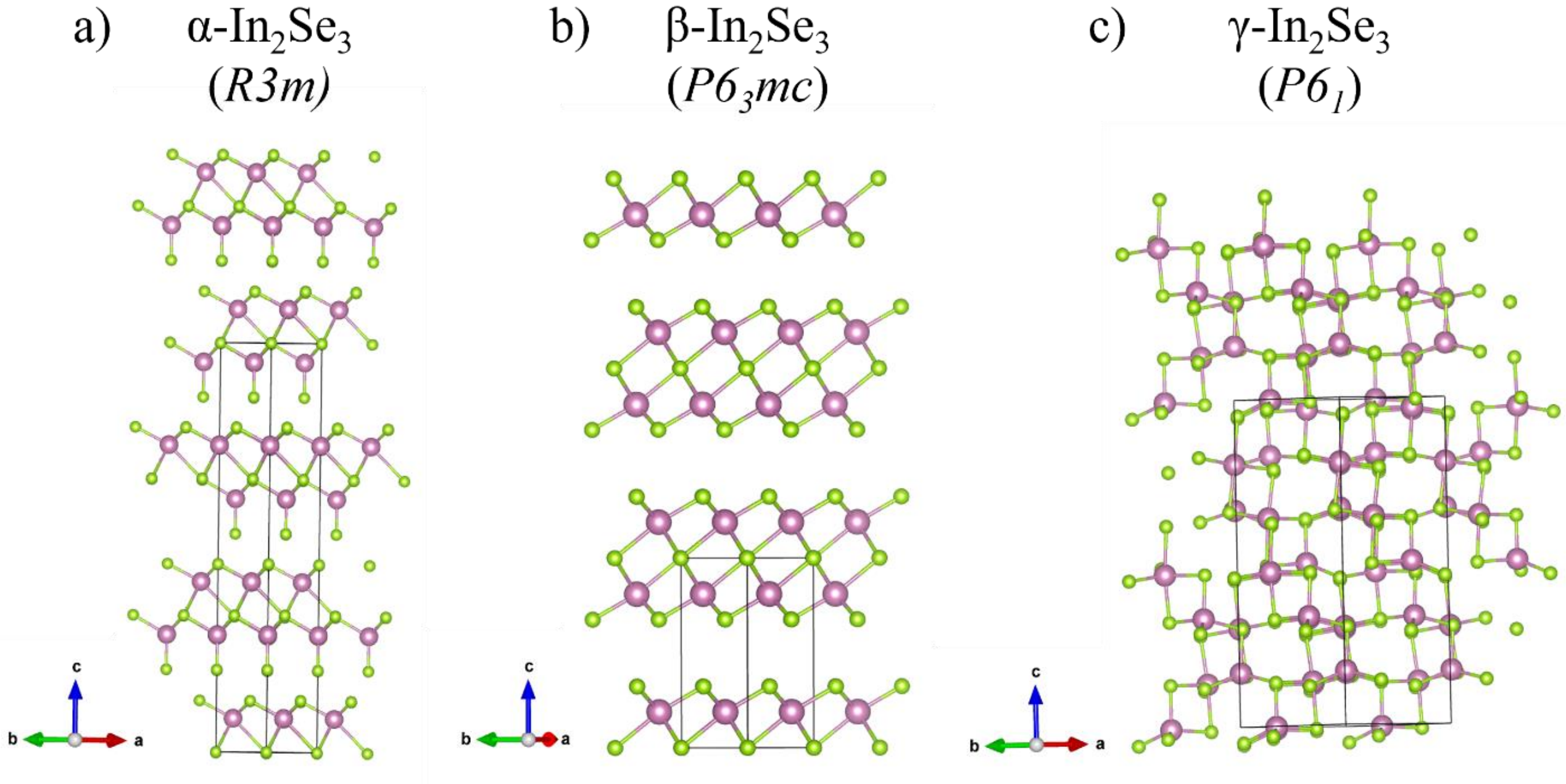

*Figure 1:Unit cell views of the common $In_2Se_3$ polymorphs a) α, b) β, and c) γ. The space groups are labeled below each polymorph.*

In this study, we apply BO to efficiently explore the growth space for $In_2Se_3$ thin films on $Al_2O_3$ substrates using MBE. We start by targeting the growth of polymorph pure γ-$In_2Se_3$. By integrating prior experimental data and employing GPR, we demonstrated a data-driven synthesis campaign that significantly reduces the number of required experiments and results in films of polymorph-pure γ-$In_2Se_3$. To further understand the influence of individual growth parameters on model predictions, we used SHapley Additive exPlanations (SHAP). SHAP uses a game-theoretic approach to explain the output of ML models and can be used to provide insights into how the ML models treat different growth conditions. Additionally, we attempt to isolate the α-polymorph and discuss the broader implications for phase-selective growth strategies in complex material systems.

## II. EXPERIMENTAL

### A. $In_2Se_3$ Synthesis using MBE

Thin films of $In_2Se_3$ were grown on (10 ×10) mm single crystal substrates of single side polished $Al_2O_3$ (0001) from Cryscore Optoelectronic Limited. Before loading, the substrates underwent sequential 10-min-long ultrasonic baths in acetone, isopropyl alcohol, and de-ionized (DI) water at room temperature. Following this the substrates were submerged in a bath of nanostrip for 40 min at a temperature of 140°C. Next, the samples were further sonicated in DI water for 10 minutes and dried with nitrogen gas. The $Al_2O_3$ wafers were then transferred into the MBE load lock and outgassed overnight at 200°C.

Upon transfer into the MBE chamber (DCA Instruments R450, DOI 10.60551/gqq8-yj90) the $Al_2O_3$ substrates were heated to 800 °C at a rate of 20 °C/min and held there for 10 minutes. The substrates were then cooled to their growth temperature (between 150°C and 550°C) with a cooling rate of 50 °C/min. $In_2Se_3$ thin films were grown by co-supplying In and Se. Indium was supplied from a dual filament Knudsen effusion cell, and selenium was supplied from a Veeco Instruments cracker source[58]. A quartz crystal microbalance (QCM – Colnatec) inserted at the sample position was used to calibrate both In and Se fluxes prior to growth to grow 20 nm thick films. After Phase 1, each selenium flux was measured at 900°C with the QCM. The QCM tooling factors for both elements were determined independently by physical film thickness measurements. Substrate temperature during deposition ranged from 150 to 550 °C as measured at the back of the wafer by a non-contact thermocouple (TC).

### B. *XRD, Raman, and AFM*

X-ray diffraction (XRD) was carried out ex situ using CuKα1 radiation in a high-resolution Panalytical X'Pert3 four-circle diffractometer with a PIXcel 3D detector. A hybrid Ge(220) crystal monochromator with 1/32° slit and 10 mm mask was used as the X-ray beam optics.

A Horiba LabRAM HR Evolution spectrometer system with unpolarized 532 nm laser excitation and 14 mW power was used to perform Raman spectroscopy measurements. Spectra were collected over 100s seconds in backscattering geometry through a 50× objective. A grating of 1800 g $mm^{-1}$ ensured the highest possible spectral resolution.

### C. *Data Preparation & Visualization*

Raman data were fitted using a Voigt profile from the SciPy package. While every peak was fitted, only the main peaks of 104 $cm^{-1}$ for α-$In_2Se_3$, 110 $cm^{-1}$ for β-$In_2Se_3$, 150 cm-1 for γ-$In_2Se_3$, and 115 $cm^{-1}$ for InSe were used to calculate the fraction of each polymorph and phase to an accuracy of 2% based on the peak fitting software. These fits can be seen in the supplementary material with Figures S5-S10 for Phase I, Figures S12-S14 for Phase II, Figure S16 for Phase III, Figure S19 for Phase IV, and Figures S27 and S28 for Phase V.

The package Bayesian Optimization by Fernando Nogueira was used for performing the Bayesian optimization[59]. This package uses Gaussian Process Regression (GPR) with the Matern 2.5 kernel trained on the observed combination of parameter and their associated target values. To suggest the next point to explore, we used the built-in Upper Confidence Bound acquisition function with κ = 0.1. The Upper Confidence Bound acquisition function follows the following equation:

$$UBC(x) = \mu(x) + \kappa\sigma(x) \quad (1)$$

where μ is the predicted mean of the GPR surrogate model, σ is the predicted standard deviation, and κ helps to control the trade-off between exploration and exploitation. A lower κ dictates more exploitation of the known parameter space, often leading to a more aggressive optimization path. The samples were done in batches of 2 or 3 following an ask-tell set up and a Kriging Believer strategy (i.e the previously predicted points are assumed to be true, and the next point is predicted off that added assumption).

To better visualize the effects of two different growth variables, color plots were used to show how the GPR was predicting our target features. The color plots shown in Section II follow a partial dependence equation for two variables while marginalizing over n other varying features. This is seen in the Equation (2).

$$f_s(x_s{}^1, x_s{}^2) = (1/n) * \Sigma_{i=1}{}^n f(x_s{}^1, x_s{}^2, x_C^{(i)}) \quad (2)$$

in which $\hat{f}$ is the prediction function of the model, $x_s{}^1$ and $x_s{}^2$ are the two features of interest, and $x_C^{(i)}$ is the values of the remaining n features for the $i^{th}$ instance in the dataset. This is useful for understanding the joint effect of two features on the predicted outcome of a machine learning model. It is important to note each $(x_s{}^1, x_s{}^2)$ coordinate point shows an average of all possible combinations of the $x_C^{(i)}$ values. This means that the color plots are better suited to showing general trends in the multidimensional growth spaces vs showing the actual maximum and minimum points.  For developing the partial dependence plots (PDPs), the same equation was used with a single specified feature of interest, $x_s{}^1$. This can be seen in Equation (3) .

$$f_s(x_s{}^1) = (1/n) * \Sigma_{i=1}{}^n f(x_s{}^1, x_C^{(i)}) \quad (3)$$

SHapley Additive exPlanations (SHAP) was used to better understand how each feature affects the GPR prediction of the target features. We used the data package SHAP, based on research by Su-In Lee[60].

## III. RESULTS AND DISCUSSION

To use BO to control the synthesis of $In_2Se_3$, we first must determine which growth parameters would be controlled and used as model inputs. There are a wide range parameters that could be varied including substrate material, growth methods (such as co-deposition or metal modulated epitaxy), synthesis parameters, annealing parameters, and so forth. For this study, we used a single substrate material (c-plane $Al_2O_3$) and co-deposition of selenium and indium with no anneal. We varied four growth parameters: substrate growth temperature ($T_G$), indium flux ($F_{In}$), selenium cracker temperature ($T_{Se}$), and selenium flux ($F_{Se}$). For this experiment, we primarily used a valved selenium cracking source with a variable temperature cracking zone. The temperature of the cracking zone determines the composition of the selenium vapor. Lower selenium cracker temperatures produce more large selenium molecules while higher cracking zone temperatures produce more reactive monomers and dimers[58,61–64]. Further details on the growth can be found in Section II.

### A. *Initial Exploration of $In_2Se_3$ on $Al_2O_3$*

To begin the experiment, we mined previous data on $In_2Se_3$ films that are publicly available in the Lifetime Sample Tracking (LiST) data management system hosted by the Two Dimensional Crystal Consortium Materials Innovation Platform (2DCC-MIP) at Pennsylvania State University. These samples were grown using the same MBE reactor with a constant indium flux and selenium cracker temperature, changing only the substrate temperature and selenium flux. This set of growth conditions achieved films of pure β-$In_2Se_3$ on $Al_2O_3$[49] as well as mixed films. In this initial sample set, a regular Knudsen effusion cell was used for selenium rather than a selenium cracker. To incorporate these samples into the model, we set the selenium cracker temperature to 500°C and reduced the fluxes to 25% of the original value. A selenium cracker temperature of 500°C is the lowest recommended cracking zone temperature for our selenium cell and this temperature for the cracking zone should produce a selenium vapor with a similar composition to a standard Knudsen effusion cell. The flux reduction is consistent with other groups using uncracked selenium, where higher flux ratios are needed to compensate for the lower reaction rate of the uncracked selenium[61,62,65].

Six new samples were grown to serve as a comparison between the regular Knudsen effusion cell and the selenium cracker. The growth conditions of the new samples were identical to six of the original samples, and $T_{Se}$ was set to 500°C. For these samples and the samples grown during this study (Phase II through Phase V) the selenium flux was measured by QCM at a cracking zone temperature of 900°C for each sample before the reduction to the stated cracking zone temperature. This was done to provide more consistent flux measurements as the QCM reading changes based on the cracking zone temperature as seen in Figure S1 in the supplementary material. Further discussion of the selenium flux measurement can be found in Section II and the supplementary material. The Raman spectra of all Phase I samples (original data plus the six new comparator samples) were modeled to extract the fraction of γ-$In_2Se_3$ in each sample; modeling details can be found in Section IV.C. This data, along with the growth parameters, were then fed into a GPR surrogate model.

The data and ML model for this initial Phase I data set are shown in Figure 2 a)- f). The black crosses indicate experimental data points, while the color plots correspond to the GPR surrogate model of growth parameter space. Regions that are darker purple are predicted to have a lower fraction of γ-$In_2Se_3$ while regions that are green and yellow are predicted to have a higher fraction of γ-$In_2Se_3$. Each subpanel shows the experimental data points and model predictions for one pair of growth parameters. Because each subpanel can only show the dependence of the fraction of γ-$In_2Se_3$ on two variables, the influences of the other variables are averaged. This means that the color should be taken as a trend indicating regions of parameter space that would give higher or lower fractions of γ-$In_2Se_3$ rather than as an absolute value of the fraction of γ-

$In_2Se_3$. The first column of subplots shows the substrate temperature ($T_G$) vs the indium flux ($F_{In}$). The second column shows $T_G$ vs selenium cracking zone temperature ($T_{Se}$). The third column shows $F_{In}$ vs $T_{Se}$. The fourth column shows $T_G$ vs the selenium flux ($F_{Se}$). The fifth column shows $F_{In}$ vs $F_{Se}$ and the sixth column shows $T_{Se}$ vs $F_{Se}$. Each of the rows shows a different experimental phase during which the model predicted the fraction of γ-$In_2Se_3$ which will be discussed in detail in the rest of the manuscript. Figure 3 shows partial dependence plots (PDPs) for each growth variable. These plots show how the fraction of γ-$In_2Se_3$ depends on one specific parameter, averaging over all other combinations, according to the model. For example, the first column of PDPs shows how the fraction of γ-$In_2Se_3$ depends on $T_G$. Similarly to Figure 2, each row shows PDPs for a different experimental phase. The samples and their parameters are given in Table S1 in the supplementary material.

We can see from the Phase I (top row) color plots in Figure 2 that some regions of parameter space have been relatively well explored (e.g. $T_G$ vs. $F_{Se}$) while others are essentially unexplored (e.g. $F_{In}$ vs. $T_{Se}$). The use of factorial DoE is also evident as 1D and 2D grids of parameter values. As shown in Figure S2 in the supplementary material, these unexplored regions have a very high uncertainty, indicating that the model has low confidence in the predictions of the fraction of γ-$In_2Se_3$ in these regions. The PDPs for the uncertainty are shown in Figure S3 in the supplementary material. Figure 3 a)-d) shows the PDPs for our initial Phase I data set. In Figure 3 a), we see a bimodal distribution, indicating that the model predicts two different temperature windows where the growth of γ-$In_2Se_3$ could be optimized. This result is contrary to both experience and intuition. By looking at Figure 2 a), we can see that these predictions likely arise from two data points that happen to show significantly higher γ-$In_2Se_3$ (40% and 35%), rather than a trend comprising multiple data points. The counterintuitive behavior of the $T_G$ PDP indicates that more experimental data points are needed to build a better model. In the Phase I data, the best sample has 40% γ-$In_2Se_3$ and was synthesized at $T_G = 300$°C, $T_{Se} = 500$°C, $F_{In} = 1.26 \times 10^{13}\ cm^{-2}s^{-1}$, and $F_{Se} = 1.16 \times 10^{13}\ cm^{-2}s^{-1}$. The fraction of the sample comprising γ-$In_2Se_3$ was determined by Raman spectroscopy measurements, an example of which is shown in Figures S5-S10 in the supplementary material.

To further understand the influence of individual growth parameters on model predictions, we applied the SHAP method. Details of the analysis are given in Figure S4 of the supplementary material. SHAP uses a game-theoretic approach to explain the output of ML models. It assigns each growth parameter ($T_G, F_{In}, T_{Se}, F_{Se}$) an importance value for a given prediction by computing its marginal contribution across all possible combinations of growth parameters. This method enables us to identify which growth parameters most significantly impact the predicted fraction of γ-$In_2Se_3$. For Phase I, the SHAP analysis showed a mean absolute SHAP value of 0.07 for $F_{Se}$ and a mean absolute SHAP value of 0.03 for $T_G$ (with zeros for both $F_{In}$ and $T_{Se}$ as those were not varied in Phase I). These SHAP values indicate that $F_{Se}$ is the most influential parameter for the GPR model, contributing an average of ±7% to the predicted fraction of γ-$In_2Se_3$. Looking at the data in a beeswarm plot shown in Figure S4 in the supplementary material, we can see that a higher $F_{Se}$ contributes negatively to the fraction of γ-$In_2Se_3$, while a lower $F_{Se}$ contributes positively. This is in agreement with the color plots, Figure 2 d)-f), and the PDP, Figure 3 d), for $F_{Se}$, where we see a higher predicted fraction of γ-$In_2Se_3$ when $F_{Se}$ is lower.

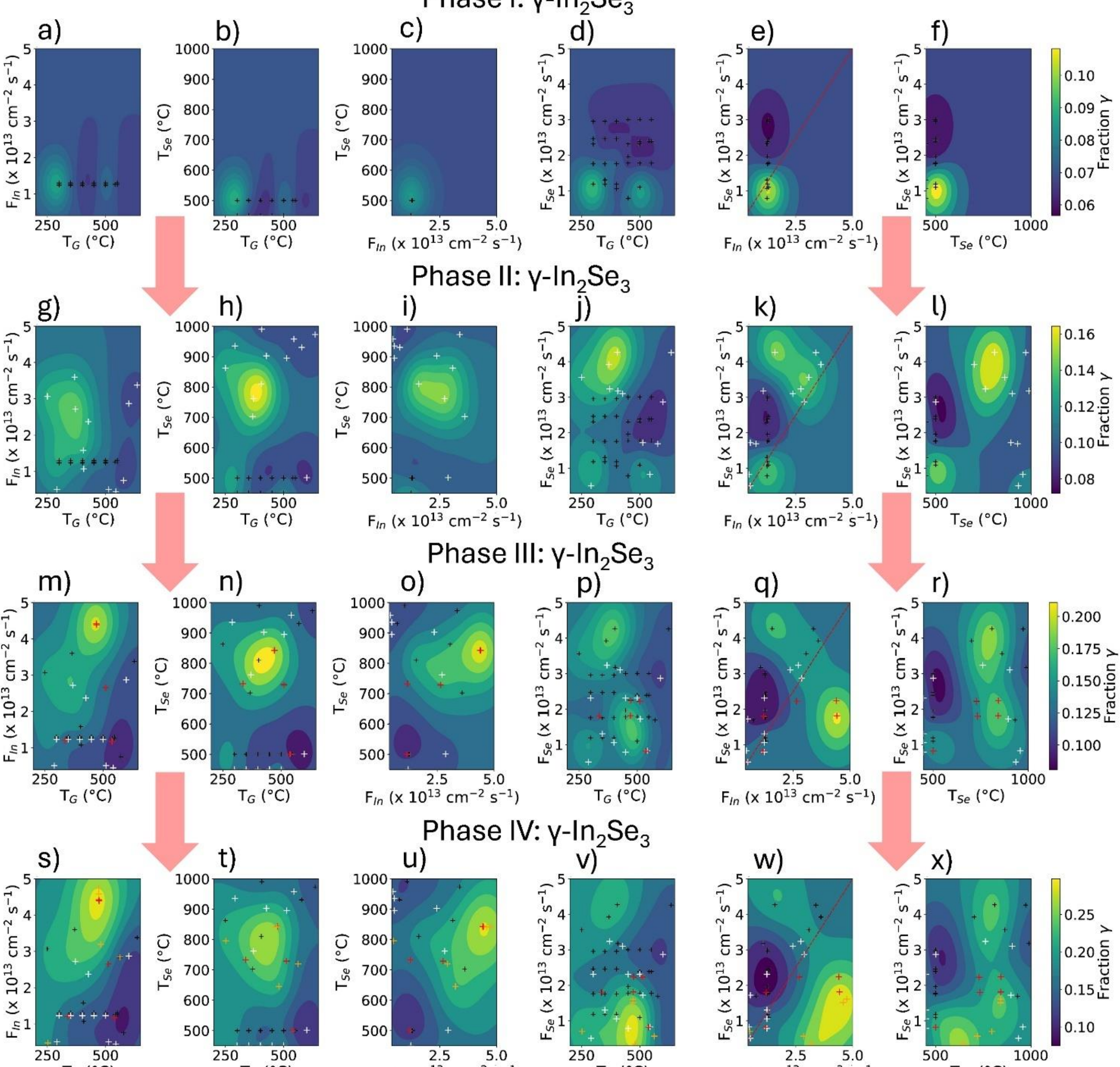

*Figure 2: Color plots for Phase I, a)-f), Phase II, g)-l), Phase III, m)-r), and Phase IV, s)-x). The black crosses indicate Phase I samples, white crosses indicate Phase II samples, red crosses indicate Phase III samples, orange crosses indicate Phase IV samples, and the color indicates the ML predicted $\gamma$-$In_2Se_3$ fraction. a), g), m), and s) show $T_G$ vs $F_{In}$. b), h), n), and t) show $T_G$ vs $T_{Se}$. c), i), o), and u) show $F_{In}$ vs $T_{Se}$. d), j), p), and v) show $T_G$ vs $F_{Se}$. e), k), q), and w) show $F_{In}$ vs $F_{Se}$. The red dashed line in the $F_{Se}$ vs $F_{In}$ plots shows the $F_{Se}/F_{In} = 1$ condition. f), l), r), and x) show $T_{Se}$ vs $F_{Se}$.*

Before we can use BO to synthesize polymorph-pure $\gamma$-$In_2Se_3$ films, we first need to further explore parameter space to improve the model. We used a Sobol sequence to generate a list of quasi-random sets of synthesis parameters. Unlike random sampling, a Sobol sequence is a low-discrepancy sequence that systematically fills the space to reduce gaps and clustering[66,67]. We restricted the parameter space to $200°C \leq T_{sub} \leq 625°C$, $7 \times 10^{12}/cm^2s \leq F_{In} \leq 6 \times 10^{13}/cm^2s$, $2 \times 10^{12}/cm^2s \leq F_{Se} \leq 4.8 \times 10^{13}/cm^2s$, and $500°C \leq T_{Se} \leq 1000°C$ based on the limitations of our system and our knowledge of what temperatures and fluxes typically result in high-quality samples. We further required $F_{Se}/F_{In} > 1$ to avoid the InSe phase. From the Sobol sequence list, we synthesized 12 samples using parameters given in Table S2 in the supplementary material; these are the Phase II samples. After growth, the fraction of γ-$In_2Se_3$ in the Phase II samples was determined by Raman spectroscopy, as discussed previously.

The Phase II samples are shown in Figure 2 g)-l) in white crosses. The original Phase I samples are shown as black crosses. The PDPs for Phase II can be seen in Figure 3 e)-h) We see that the color plots and the PDPs have changed significantly compared to the Phase I data. The uncertainty in the model has also reduced, as shown in Figure S2 in the supplementary material. The PDP of $T_G$ now shows a single peak, as expected, rather than the double peak behavior we saw after Phase I. In the PDP for $T_{Se}$, we see a peak around 750°C, giving a selenium vapor with a mix of both large molecules (uncracked) and monomers and dimers (cracked). The best Phase II sample yielded an $In_2Se_3$ film with 51%±2% γ polymorph, as measured by Raman spectroscopy.

The SHAP analysis for Phase II can be seen in Figure S11 of the supplementary material. For Phase II, the SHAP analysis showed a mean absolute SHAP value of 0.06 for $F_{Se}$, 0.04 for $T_{Se}$, 0.03 for $T_G$, and 0.01 for $F_{In}$. These SHAP values indicate that $F_{Se}$ remains the most influential parameter for the GPR model, contributing an average of ±6% to the predicted γ-$In_2Se_3$ polymorph fraction.

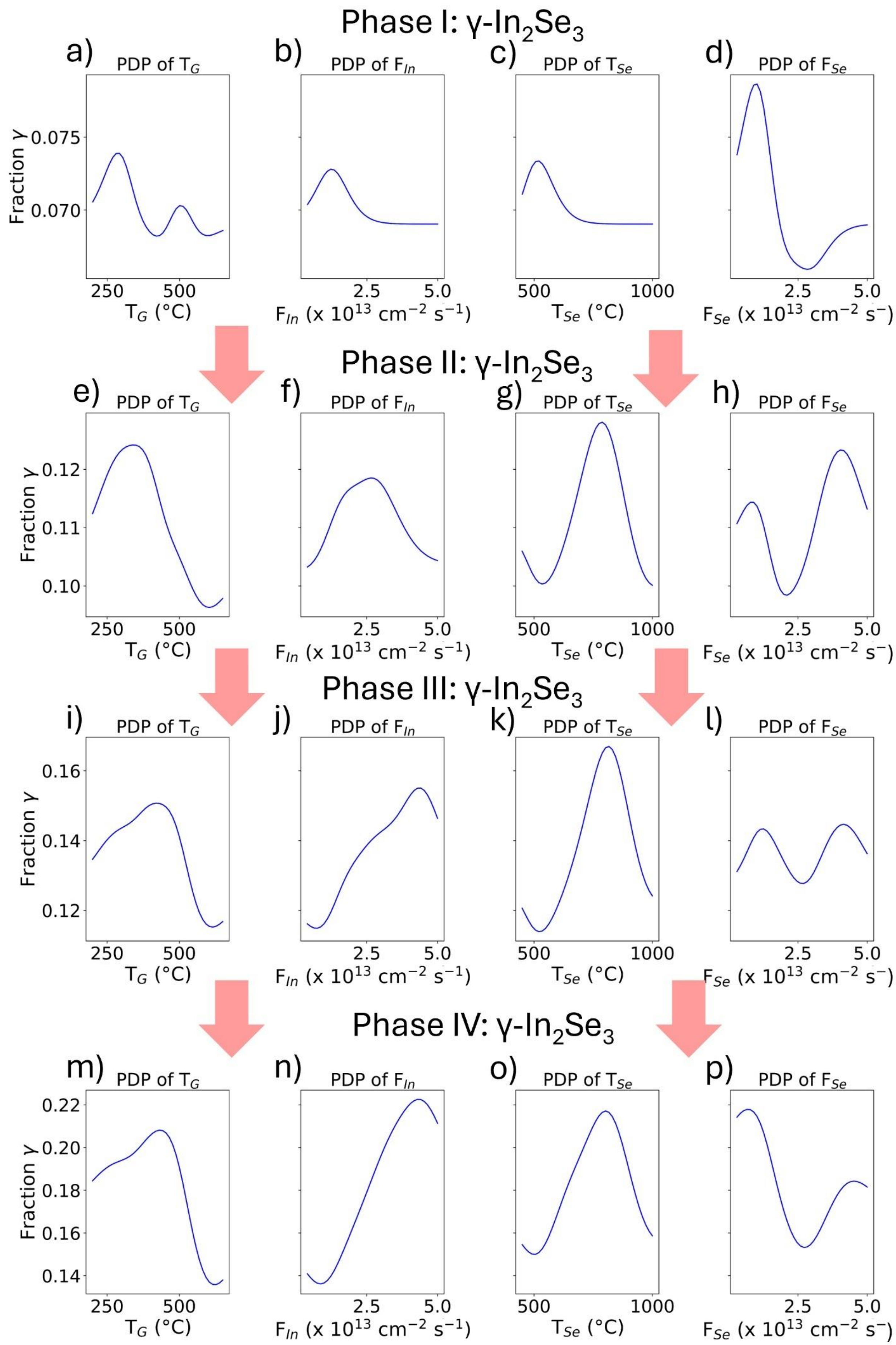

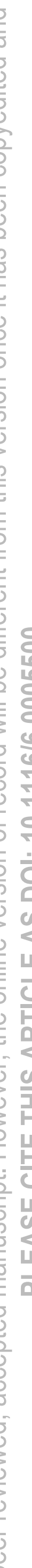

*Figure 3: Partial dependance plots (PDP) for Phase I, a)-d), Phase II, e)-h), Phase III, i)-l), and Phase IV, m)-p), showing the fraction of γ-$In_2Se_3$. a), e), i), and m) show $T_G$ vs fraction of γ-$In_2Se_3$. b), f), j), and n) show $F_{In}$ vs fraction of γ-$In_2Se_3$. c), g), k), and o) show $T_{Se}$ vs fraction of γ-$In_2Se_3$. d), h), l), and p) show $F_{Se}$ vs fraction of γ-$In_2Se_3$.*

## B. *Bayesian Optimization of γ-$In_2Se_3$ on $Al_2O_3$*

After sufficiently exploring parameter space, we focused on achieving polymorph-pure γ-$In_2Se_3$ films using BO with a GPR surrogate. Details of the BO algorithm are given in Section IV.B. Five samples were grown in Phase III; details of the growth parameters are given in Table S3 in the supplementary material. Figure 2 m)-r) shows the color plots including data from Phase I (black crosses), Phase II (white crosses), and Phase III (red crosses). Figure 3 i)-l) show the Phase III PDPs. In Phase III, the $F_{Se}/F_{In} > 1$ requirement was removed. Upon removal of this constraint, the BO algorithm targeted the $F_{Se}/F_{In} < 1$ unexplored region due to the high uncertainty in that region, which can be seen in Figure S2 in the supplementary material. In comparing Phase II and Phase III of Figure 2, we see that samples grown in this region show a high fraction of γ-$In_2Se_3$. We also see that the PDP for $F_{In}$ shifts to higher fluxes and thus higher growth rates. The first sample grown using Bayesian Optimization had 55%±2% γ-$In_2Se_3$, while the fifth sample increased to 80%±2% γ-$In_2Se_3$. This demonstrates another advantage of using an ML model, which is free from the assumption that $In_2Se_3$ needed to be grown in an $F_{Se}/F_{In} > 1$ environment.

The SHAP analysis for Phase III can be seen in Figure S15 of the supplementary material. For Phase III, the SHAP analysis showed a mean absolute SHAP value of 0.07 for $F_{In}$, 0.05 for $F_{Se}$, 0.03 for $T_G$, and 0.02 for $T_{Se}$. These SHAP values indicate that $F_{In}$ is now the most influential parameter for the GPR model, contributing an average of ±7% to the predicted γ-$In_2Se_3$ fraction. Looking at the beeswarm plot in Figure S15 in the supplementary material, we can see that the samples with high $F_{In}$ are contributing much more positively to the predicted γ-$In_2Se_3$ fraction while the low $F_{In}$ values contribute slightly negatively. In isolation, this indicates that a higher $F_{In}$ (and thus faster growth rate) is one of the primary factors for our higher γ-$In_2Se_3$ percentage growths. However, as $F_{Se}$ is comparable in magnitude, an interplay between $F_{In}$ and the $F_{Se}$ could instead be the underlying factor. Disentangling interdependent parameters using SHAP is extremely difficult and beyond the scope of this paper.

Following this successful increase in the fraction of γ-$In_2Se_3$, we chose to grow five more samples for Phase IV to determine the maximum fraction of γ-$In_2Se_3$ that could be achieved within the growth constraints. The growth parameters of the Phase IV samples are given in Table S4 in the supplementary material. Figure 2 s)-x) shows the color plots for all samples from Phase I through Phase IV; the new samples are shown as orange crosses. All five Phase IV samples were synthesized in the $F_{Se}/F_{In} < 1$ region. Figure 3 m)-p) shows the PDPs. The best sample showed 84%±2% γ-$In_2Se_3$ when measured by Raman spectroscopy and 91%±2% γ-$In_2Se_3$ when measured by x-ray diffraction, as shown in Figure S20 in the supplementary material.

This was achieved by growing less than 10 samples following the introduction of the BO scheme. While achieving this high fraction of γ-$In_2Se_3$ in 10 samples is relatively quick, it is important to note that a total of 51 samples were used, including leveraging data from 23 samples grown years before this study. BO could also have been done starting with no previous data, which may have been more efficient. However, the ability to include data from previous growth campaigns is valuable, even if those samples are not well-dispersed throughout parameter space.

SHAP analysis showing the bar plots and beeswarm plots for all samples through Phase IV can be seen in Figures 4 a) and b), respectively. Figure 4 a) shows the mean absolute SHAP value for each parameter. The bar plots are ranked vertically by relative influence of each parameter on the ML model. We see a mean absolute SHAP value of 0.13 for $F_{In}$, 0.07 for $F_{Se}$, 0.04 for $T_G$, and 0.02 for $T_{Se}$. These SHAP values indicate that $F_{In}$ is the most influential parameter for the GPR model, contributing an average of ±13% to the predicted γ-$In_2Se_3$ polymorph fraction. The beeswarm plots in Figure 4 b) show the underlying SHAP value for each sample from Phases I-IV and how they contribute to the model's predictions. Each point represents one sample. The position of the point on the x-axis shows the SHAP value of that sample for that parameter, while the color bar shows the relative raw value of the parameter for the sample. For example, the right-most point in the $F_{In}$ line in Figure 4 b) shows a SHAP value of ~0.60 and the red color indicates that that sample had a high $F_{In}$. From Figure 4 b), we can see that lower values of $F_{In}$ (blue dots) have negative SHAP values while higher values of $F_{In}$ have positive SHAP values. This indicates that the samples grown with a lower $F_{In}$ have a lower predicted fraction of γ-$In_2Se_3$ while samples grown with a higher $F_{In}$ have a higher predicted fraction of γ-$In_2Se_3$. We see the reverse trend in $F_{Se}$, where samples grown with a lower $F_{Se}$ have a positive SHAP value and samples grown with a higher $F_{Se}$ have a lower SHAP value. These correlate well with the PDPs of $F_{In}$ and $F_{Se}$ shown in Figure 3 n) and p). Additionally, further post hoc analysis through Leave-One-Out cross validation followed by drop-column feature importance aligns with showing $F_{In}$ as the most important feature. This can be seen in Figure S17 in the supplementary material. Comparison of using a GPR vs Random Forest vs Linear regression shows that a Random Forest surrogate model does perform slightly better than the GPR and Linear regression models, which can be seen in Figure S18 in the supplementary material.

It is interesting to compare the color plots and PDPs from Phase II to Phase IV. We see that the peak for $T_G$ moves from approximately 300°C to approximately 450°C. The peak for $F_{In}$ shifts to higher fluxes, while the peak for $F_{Se}$ shifts to lower fluxes, resulting in a much lower flux ratio. These shifts show that γ-$In_2Se_3$ forms with a lower Se:In flux ratio. We hypothesize that this polymorph is stabilized by selenium vacancies which form more readily at low Se:In flux ratios and high substrate temperatures. Polymorphs of other 2D materials have also been demonstrated to be stabilized by defects, making this a plausible explanation[68–71]. It is important to note that although our only goal was to maximize the fraction of γ-$In_2Se_3$ in the film, we were still able to achieve films with good crystalline quality as measured with a measured an XRD rocking curve with a full width half maximum (FWHM) of 0.10°±0.01° and a FWHM by Raman of 9.6±0.1 $cm^{-1}$. The root mean square (RMS) roughness as measured by AFM was 17 nm ± 1nm. While this film RMS roughness is likely too large for device usage, this is a parameter that could later be minimized with a multi-objective Bayesian optimization. The XRD rocking curve can be seen in Figure S21 and the AFM in Figure S22 of the supplementary material. This low FWHM is unlikely to have happened if we truly had a lower selenium than indium flux, which would produce indium droplets. Instead, we hypothesize that our measurement of $F_{Se}$ underestimates the true $F_{Se}$. We use a QCM to measure elemental fluxes, and the QCM is calibrated to the thickness of a pure elemental film deposited at room temperature. Inherent in this calibration is the assumption that 100% of the deposited material sticks to the substrate at room temperature and to the QCM itself. Given the extreme volatility of selenium, it may be the case that we have significant re-evaporation both when depositing the thickness standard and during daily QCM measurements. This could lead to selenium fluxes that are reproducible, but that underestimate the real flux, resulting in a systematic error in our flux ratios. It is important to note that without a data-driven approach, it is highly unlikely that we would have explored the $F_{Se}/F_{In} < 1$ region and we would not have achieved polymorph-pure γ-$In_2Se_3$ films. Furthermore, the GPR is trained on control parameters rather than the true state of the growth chamber; since we cannot know the exact chemical environment in the chamber, our intuition-based approach will most likely introduce bias. These factors clearly demonstrate the benefit of using ML for growth optimization.

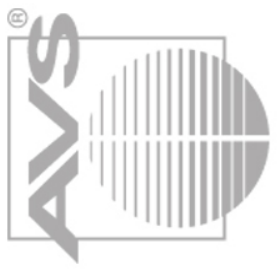

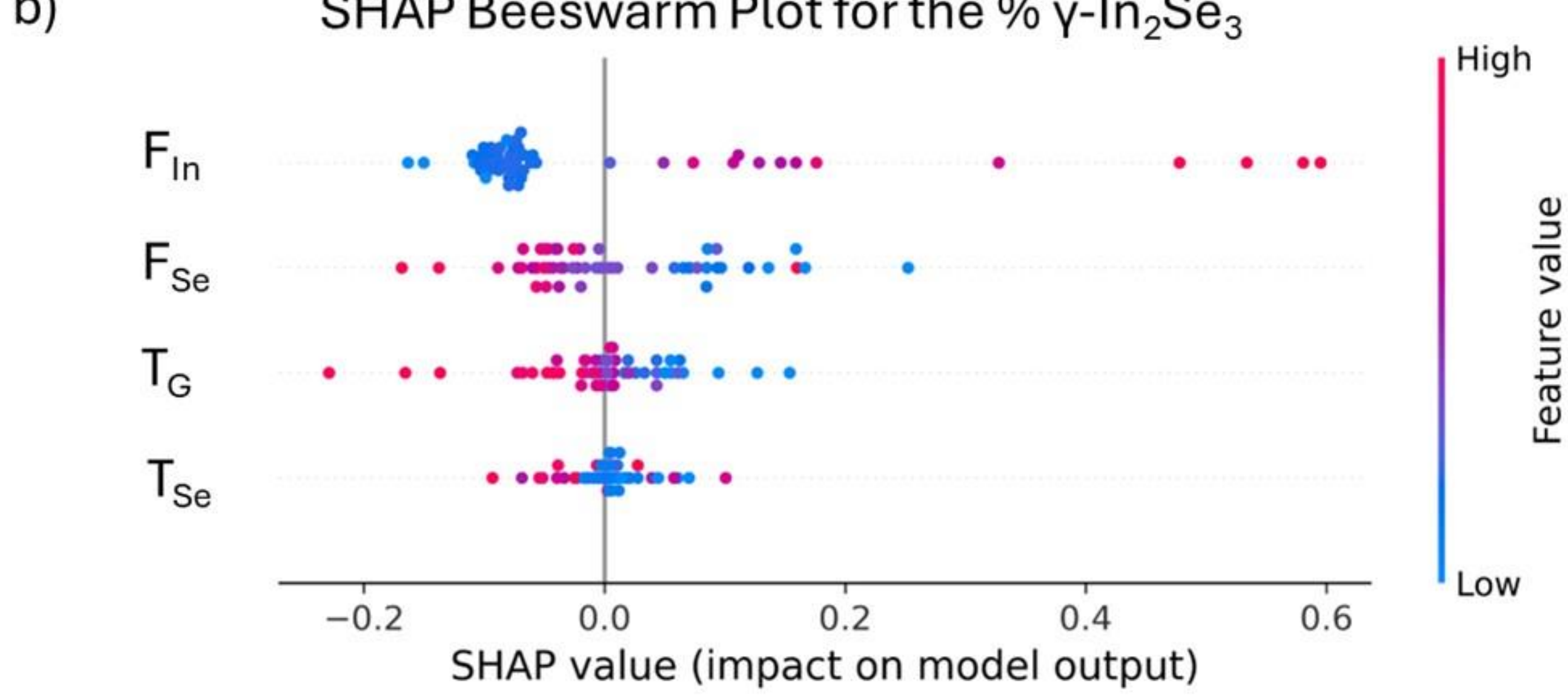

*Figure 4: a) Bar plot showing the mean absolute SHAP value for each parameter using data from Phases I-IV. The bar plots are ranked by relative influence of each parameter to the model's predicted fraction $\gamma$-$In_2Se_3$. b) Beeswarm plot ranked by mean absolute SHAP value using data from Phases I-IV. Each sample appears as its own point and are distributed along the x-axis according to their SHAP value. Areas with a higher density of SHAP values are stacked vertically while the color bar corresponds to the raw parameter value for each sample.*

### C. *Bayesian Optimization of α-$In_2Se_3$ on $Al_2O_3$*

Following the successful synthesis of γ-$In_2Se_3$ films, we attempted to use BO to grow pure α-$In_2Se_3$ films. The initial color plots are shown in Figure 5 a)-f) using the samples from Phases I-IV discussed earlier. The PDP plots using samples from Phases I-IV are shown in Figure 6 a)-d). In Figure 5, the black crosses indicate experimental data points from Phase I and II. The white crosses indicated experimental data points from Phase III and IV. The red crosses indicate experimental data points from Phase V, while the color plots correspond to the GPR surrogate model of growth parameter space. Regions that are darker purple are predicted to have a lower fraction of α-$In_2Se_3$ while regions that are green and yellow are predicted to have a higher fraction of α-$In_2Se_3$. Each subpanel shows the experimental data points and model predictions for one pair of growth parameters. Because each subpanel can only show the dependence of the fraction of α-$In_2Se_3$ on two variables, the influences of the other variables are averaged. Figure 6 shows partial dependence plots (PDPs) for each growth variable. These plots show how the fraction of α-$In_2Se_3$ depends on one specific parameter, averaging over all other combinations, according to the model. For example, the first column of PDPs shows how the fraction of α-$In_2Se_3$ depends on $T_G$. Similarly to Figure 5, each row shows PDPs for a different experimental phase. From Figure 6, we can clearly see that the fraction of α-$In_2Se_3$ is strongly dependent on $T_G$ and shows a peak around 250°C. This is unsurprising, given that $In_2Se_3$ undergoes a phase transition from α to β around 200°C - 300°C[31,32]. The figures showing the uncertainty color map and the PDPs for α-$In_2Se_3$ are shown in Figure S23 and S24, respectively, in the supplementary material.

This temperature dependence correlates well with the SHAP analysis for α-$In_2Se_3$, which can be seen on Figure S25 in the supplementary material. The SHAP analysis showed a mean absolute SHAP value of 0.03 for $T_G$, 0.01 for $T_{Se}$, 0.01 for $F_{Se}$, and 0.00 for $F_{In}$. These SHAP values indicate that $T_G$ is the most influential parameter for the GPR model. Looking at the beeswarm plot, we see that lower temperature growths have a higher positive SHAP value, while higher temperature growths show a negative SHAP value, indicating that lower $T_G$ correspond with higher fractions of α-$In_2Se_3$. However, as none of the films contain a significant percentage of α-$In_2Se_3$ the SHAP values are very low. This trend fits well with our knowledge of the α-$In_2Se_3$ phase transition.

*Figure 5: Color plots for Phase IV, a)-f), and Phase V, g)-l), showing the fraction of α-$In_2Se_3$. The black crosses indicate Phase I and II samples, white crosses indicate Phase III and IV samples, red crosses indicate Phase V samples, and the color indicates the ML predicted α-$In_2Se_3$ fraction. a) and g) show $T_G$ vs $F_{In}$. b) and h) show $T_G$ vs $T_{Se}$. c) and i), show $F_{In}$ vs $T_{Se}$. d) and j) show $T_G$ vs $F_{Se}$. e) and k) show $F_{In}$ vs $F_{Se}$. The red dashed line in the $F_{Se}$ vs $F_{In}$ plots shows the $F_{Se}/F_{In} = 1$ condition. f) and l) show $T_{Se}$ vs $F_{Se}$.*

To optimize for polymorph-pure α-$In_2Se_3$, we used BO to suggest the next nine samples, as described above. Figure 5 g)-l) shows these new samples in red crosses, as well as the color plots and PDPs for the ML model of α-$In_2Se_3$. Samples from Phases I and II are shown in black crosses and samples from Phases III and IV are shown as white crosses. The new samples and their parameters given in Table S5 in the supplementary material. The BO algorithm primarily suggested samples where $T_G < 300$°C, with a few samples at higher $T_G$ to help fill out the parameter space. Focusing on low $T_G$ aligns with our intuition about the best growth parameters for α-$In_2Se_3$, given the existence of the structural transition. Unfortunately, these new samples showed no increase in the fraction of α-$In_2Se_3$.

This temperature dependence again correlates well with the SHAP analysis for α-$In_2Se_3$, which can be seen in Figure S26 in the supplementary material. The SHAP analysis showed a mean absolute SHAP value of 0.03 for $T_G$, 0.01 for $T_{Se}$, 0.01 for $F_{Se}$, and 0.01 for $F_{In}$. These SHAP values still indicate that $T_G$ is still the most influential parameter for the GPR model.

The maximum fraction of α-$In_2Se_3$ achieved using BO was 27%±2%, which is only slightly better than the previous maximum of 22%±2%. We attribute this failure to the fact that low growth temperatures are needed to stabilize α-$In_2Se_3$, but at low growth temperatures, the films became amorphous or polycrystalline. Because we want single crystalline films, we penalized any samples that were amorphous or polycrystalline by assigning them values of 0% in all polymorphs and phases. This experiment leads us to determine that it is extremely challenging if not impossible to grow crystalline α-$In_2Se_3$ on $Al_2O_3$ using 1-step growth and co-deposition within this growth parameter window. Thus, even in the case of negative results, BO increases the efficiency of our experimental campaign, avoiding wasted time and effort in an intuition-based trial-and-error approach.

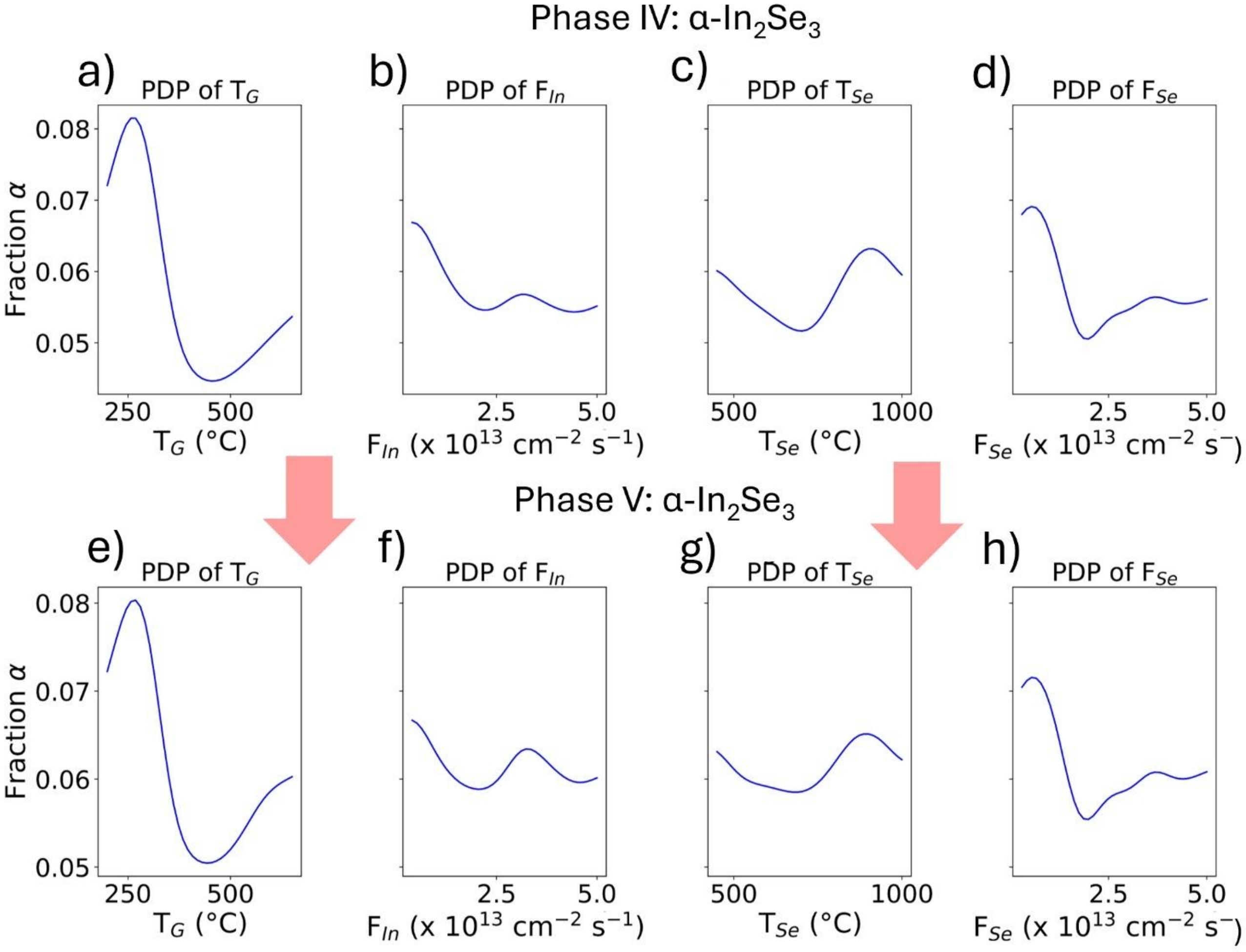

*Figure 6: Partial dependance plots (PDP) for Phase IV, a)-d), and Phase V, e)-h), showing the fraction of α-$In_2Se_3$. a) and e) show $T_G$ vs fraction of α-$In_2Se_3$. b) and f) show $F_{In}$ vs fraction of α-$In_2Se_3$. c) and g) show $T_{Se}$ vs fraction of α-$In_2Se_3$. d) and h) show $F_{Se}$ vs fraction of α-$In_2Se_3$.*

## IV. Conclusion

In this work, we demonstrated an effective data-driven campaign using Bayesian Optimization to guide the growth of polymorph-pure γ-$In_2Se_3$ thin films on $Al_2O_3$ substrates using MBE. BO guided us towards a Se:In flux ratio below 1:1, a region that we would not have otherwise explored without surrogate modeling, highlighting the usefulness of data-driven models in experimental design. In addition, leveraging BO resulted in an increase in the fraction of γ-$In_2Se_3$ from 51% to 84% in fewer than 10 samples. Additional work on the γ-$In_2Se_3$ polymorph could benefit from a multi-objective BO focused on maintaining a high fraction of γ-$In_2Se_3$ while reducing the RMS roughness of the film. Attempts to isolate the α-polymorph using BO were limited by the formation of amorphous or polycrystalline films at lower temperatures; we believe this suggests that single-step co-deposition on $Al_2O_3$ may not be viable for crystalline α-$In_2Se_3$, rather than the case that we did not find a precise parameter set to achieve the desired outcome. Further expansion of BO using categorical values for different growth procedures could allow for further progress. Other growth procedures that could be used to help better isolate the α-polymorph could range from metal modulated epitaxy to a simple two-step growth sequence with deposition at different substrate temperatures. Additionally, adding data for α-$In_2Se_3$ grown on various substrates might help in achieving crystalline α-$In_2Se_3$ through MBE growth. These findings highlight the potential of data-driven optimization in complex materials synthesis and further solidify the growing body of evidence supporting the use of ML for efficient thin film experimental design.

## Supplementary Material

The supplementary material includes the QCM measurements of the selenium flux at different cracking zone temperatures. Included are the color plots and partial dependence plots for the GPR calculated standard deviation for Phase I-IV looking at γ-$In_2Se_3$, and Phase IV and V looking at α -$In_2Se_3$. For Phase I, II, and III looking at γ-$In_2Se_3$ the SHAP analysis is included along with the SHAP analysis for Phase IV and V looking at α -$In_2Se_3$. The supplementary material also contains tables showing the growth conditions and polymorph fractions for each sample. Additionally, the fitted Raman plots are included as well. XRD, AFM, and rocking curve images are included for sample 48, the highest fraction of γ-$In_2Se_3$ sample. Graphs looking at drop-column feature importance and the performance of different regression models for Phase IV are also added here.

## Acknowledgements

This study is based on research conducted at the Pennsylvania State University Two-Dimensional Crystal Consortium—Materials Innovation Platform (2DCC-MIP), which is supported by the NSF cooperative agreement DMR-2039351. The authors acknowledge funding from the Pennsylvania State University Materials Research Institute Seed Grant Program. M. A. acknowledges support from the National Science Foundation Grant No. EEC-2244201.

## Conflicts of Interest

The authors declare no conflict of interest.

## Data availability statement

All data of this study is available under the following link for the review process, which will be converted into an open-access link to the data with separate DOI in ScholarSphere upon publication: https://data.2dccmip.org/g1nk5cdZV2io

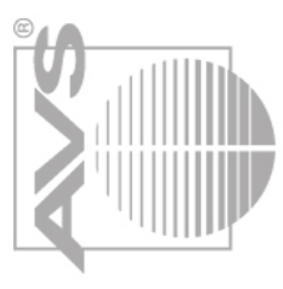